\documentclass[12pt]{article}

\usepackage{amsmath}
\usepackage{epsfig}
\usepackage{amssymb}
\usepackage[colorlinks=true,
            linkcolor=blue,
            citecolor=blue]{hyperref}
\usepackage{bbold}
\usepackage{braket}
\input{epsf}
\usepackage[english]{babel}
\usepackage{siunitx}
\usepackage[numbers,sort&compress]{natbib}

\newcommand*{\braa}[1]{\left(#1\right)}

\newcommand*{\brb}[1]{\left[#1\right]}

\newcommand*{\abs}[1]{\left|#1\right|}
\newcommand*{\fewz}{\textsc{fewz}\ }
\newcommand*{\dynnlo}{\textsc{dynnlo}\ }

\def\to{\rightarrow}

\def\nn{\nonumber}
\def\beq{\begin{equation}}
\def\eeq{\end{equation}}
\def\beeq{\begin{eqnarray}}
\def\eeeq{\end{eqnarray}}
\def\ba{\begin{eqnarray}}
\def\ea{\end{eqnarray}}
\def\beal{\begin{align}}
\def\eeal{\end{align}}

\begin{document}

\begin{titlepage}
\renewcommand{\thefootnote}{\fnsymbol{footnote}}
\par \vspace{10mm}

\begin{center}
\vspace*{10mm}
{\Large \bf
Drell-Yan lepton angular distributions \\[3mm] 
in perturbative QCD}
\end{center}

\par \vspace{2mm}
\begin{center}
{\bf Martin Lambertsen,}
\hskip .1cm
{\bf Werner Vogelsang}\\[5mm]
\vspace{5mm}
Institute for Theoretical Physics, T\"ubingen University, 
Auf der Morgenstelle 14, \\ 72076 T\"ubingen, Germany\\[2mm]
\end{center}


\vspace{9mm}
\begin{center} {\large \bf Abstract} \end{center}
We present a comprehensive comparison of the available experimental data for the Drell-Yan lepton
angular coefficients $\lambda$ and $\nu$ to calculations at leading and next-to-leading order of 
perturbative QCD. To obtain the next-to-leading order corrections, we make use of publicly available numerical codes
that allow us to compute the Drell-Yan cross section at second order in perturbation theory and from which 
the contributions we need can be extracted. Our comparisons reveal that perturbative QCD is able 
to describe the experimental data overall rather well, especially at colliders, but also in the fixed-target
regime. On the basis of the angular coefficients alone, there appears to be little (if any) convincing
evidence for effects that go beyond fixed-order collinear factorized perturbation theory, although the
presence of such effects is not ruled out.

\end{titlepage}  

\setcounter{footnote}{2}
\renewcommand{\thefootnote}{\fnsymbol{footnote}}


\section{Introduction}

It has been known for a long time~\cite{Collins:1977iv,Lam:1978pu} that leptons produced in the Drell-Yan process 
$H_1H_2\to \ell\bar\ell X$ may show nontrivial angular distributions. We denote the momentum of the intermediate 
virtual boson $V=\gamma^*,Z$ that decays into the lepton pair by $q$. In a specific rest frame of the virtual boson (for our purposes,
the Collins-Soper frame~\cite{Collins:1977iv}) we can define polar and azimuthal lepton decay angles $\theta$ and $\phi$, 
respectively. Considering, for simplicity, a situation where contributions by $Z$-bosons are negligible and only the 
exchange of an intermediate virtual photon is relevant, one can show that the cross section differential in 
$d^4q$ and $d\Omega\equiv d\cos\theta d\phi$ may be written as
\beq
\frac{d\sigma}{d^4 q d\Omega} \,=\, \frac{\alpha^2}{2\pi N_c Q^2 s^2}
	\left( W_{T}\braa{1+\cos^2\theta}+W_{L}\braa{1-\cos^2\theta}
	+ W_{\Delta}\sin2\theta\cos\phi + W_{\Delta\Delta}\sin^2\theta\cos 2\phi \right) \label{eqn: diff cross section}\,,
\eeq
where $\alpha$ is the fine structure constant, $N_c=3$ the number of colors in QCD, $Q^2=q^2$ and $s$ the
c.m.s. energy squared of the incoming hadrons $H_1$ and $H_2$. The structure functions $W_T$, $W_L$, $W_\Delta$, $W_{\Delta\Delta}$ are functions of $q$. They parametrize the hadronic tensor 
as 
\beq
{W}^{\mu \nu} = - (g^{\mu \nu}- T^\mu T^\nu) (W_T+ W_{\Delta
\Delta}) - 2 X^\mu X^\nu W_{\Delta \Delta} + Z^\mu
Z^\nu (W_L -W_T - W_{\Delta \Delta} ) - (X^\mu Z^\nu +
Z^\mu X^\nu) W_\Delta \ ,
\label{WTLDDD}
\eeq
where $X$, $Y$, $Z$ and $T$ are a set of orthonormal axes that one introduces in the Collins-Soper frame.
If also $Z$-bosons contribute, there are additional angular terms and structure functions in the cross section formula.
For details of the derivation of the cross section (also for discussion of other related reference frames), see 
Refs.~\cite{Collins:1977iv,Lam:1978pu,Boer:2006eq,Berger:2007si,Peng:2014hta}. 

From the differential cross section one easily derives an expression for the normalized decay 
angle distribution
\beq
\frac{dN}{d\Omega} \equiv \left(\frac{d \sigma}{d^4 q}
\right)^{-1} \frac{d \sigma}{d\Omega d^4 q}  \label{angdef}
\eeq
in terms of the structure functions. Using Eq.~(\ref{eqn: diff cross section}) we obtain
\ba
\frac{dN}{d\Omega}  =
\frac{3}{8\pi} \; \frac{W_T (1+ \cos^2 \theta)   + W_L (1- \cos^2\theta)
+ W_\Delta \sin 2\theta \cos \phi + W_{\Delta \Delta} \sin^2 \theta
\cos 2\phi}{2 W_T + W_L} . \label{sigmaW}
\ea
One usually writes this as 
\ba
\frac{dN}{d\Omega} & = &
\frac{3}{4\pi} \; \frac{1}{\lambda+3} \left[ 1+ \lambda \cos^2\theta
+ \mu \sin 2\theta \cos \phi + \frac{\nu}{2} \sin^2 \theta
\cos 2\phi \right],
\label{DNdOmega}
\ea
where
\ba
\lambda=\frac{W_T-W_L}{W_T+W_L}\; , \;\;\;
\mu=\frac{W_\Delta}{W_T+W_L}\; , \;\;\;
\nu=\frac{2W_{\Delta\Delta}}{W_T+W_L} \; .
\label{lmnWrel}
\ea
Much effort has gone into studies of these angular coefficients $\lambda,\mu,\nu$, both 
experimentally and theoretically. On the experimental side, measurements of the coefficients 
are by now available over a wide range of kinematics, from fixed-target 
energies~\cite{Guanziroli:1987rp,Conway:1989fs,Zhu:2006gx,Zhu:2008sj} 
all the way to the Tevatron~\cite{Aaltonen:2011nr} $p\bar{p}$ and the LHC $pp$ 
colliders~\cite{Khachatryan:2015paa}. In the fixed-target regime various combinations
of beams and targets are available; data have been taken with pion beams off nuclear (tungsten)
targets~\cite{Guanziroli:1987rp,Conway:1989fs} and also for $pp$ and $pd$ 
collisions~\cite{Zhu:2006gx,Zhu:2008sj}. 
The experimental results are typically given as functions of the transverse momentum $q_T$ 
of the virtual boson, in a certain range of the lepton pair mass, $Q\equiv\sqrt{Q^2}$. For the 
fixed-target data, $q_T$ is limited to a few GeV and $Q$ is usually around
\SIrange{5}{10}{GeV}. This is very different for the high-energy collider measurements which are carried
out around $Q=m_Z$, where $m_Z$ is the $Z$-boson mass. The range in $q_T$ explored
here is much larger and reaches to almost \SI{100}{GeV} at the Tevatron and even much beyond
that at the LHC.

The lowest-order (LO) partonic channel $q\bar{q}\to V\,(\to \ell\bar\ell)$ with collinear
incoming partons leads to the prediction $\lambda=1$, $\mu=\nu=0$. However, for this
process the virtual photon has vanishing transverse momentum, $q_T=0$, so it
cannot contribute to the cross section at finite $q_T$. The situation changes
when ``intrinsic'' parton transverse momenta are taken into account. The coefficient $\nu$, especially, which corresponds to a $\cos2\phi$ dependence in azimuthal
angle, has received a lot of attention in this context since it was discovered~\cite{Boer:1999mm} 
that it may probe interesting novel parton distribution functions of the nucleon, known
as Boer-Mulders functions~\cite{Boer:1997nt}. These functions represent a transverse-polarization 
asymmetry of quarks inside an unpolarized hadron and are ``T-odd'' and hence related to 
nontrivial (re)scattering effects in QCD (see~\cite{Collins:2002kn}). Detailed 
phenomenological~\cite{Barone:2010gk,Lu:2011mz} or model-based~\cite{Pasquini:2014ppa} studies 
have been presented that confront the fixed-target experimental data with theoretical expectations
based on the Boer-Mulders functions.

Already the early theoretical studies~\cite{Lam:1978zr,Collins:1978yt,Kajantie-78,Cleymans-78,Lindfors-79} 
revealed that also plain perturbative-QCD radiative effects lead to departures from the simple prediction
$\lambda=1$, $\mu=\nu=0$, starting from ${\cal O}(\alpha_s)$ with the processes $q\bar{q}\to Vg$ and
$qg\to Vq$. At $q_T\neq 0$ in fact the latter processes become the LO ones. A venerable result 
of~\cite{Lam:1978pu,Lam:1980uc} obtained on the basis of these LO reactions is the {\it Lam-Tung relation},
\beq\label{LT}
1-\lambda-2\nu\,=\,0\,,
\eeq
which holds separately for both partonic channels in the Collins-Soper frame~\cite{Collins:1977iv}. 
Next-to-leading order (NLO) corrections to the cross sections relevant for the angular coefficients have first been
derived in Refs.~\cite{Mirkes:1992hu,Mirkes:1994dp}. These suggest overall modest ${\cal O}(\alpha_s^2)$ 
effects on $\lambda$, $\mu$, $\nu$, so that also the Lam-Tung relation, although found to be violated at NLO, 
still holds to fairly good approximation. The data from the fixed-target experiment E615~\cite{Guanziroli:1987rp}
indicate a violation of the Lam-Tung relation, while the other fixed-target
sets are overall consistent with it, as are the Tevatron data~\cite{Aaltonen:2011nr}.
A clear violation of the Lam-Tung relation, on the other hand, was observed
recently at the highest energies, in $pp$ collisions at the LHC~\cite{Khachatryan:2015paa}.

In the present paper, we take a fresh look at the Drell-Yan angular dependences in the framework 
of perturbative QCD. Specifically, we present an exhaustive comparison of the LO and NLO
QCD predictions for the parameters $\lambda$ and $\nu$ with the experimental data, over the 
whole energy range available. Rather than attempting to retrieve the results of~\cite{Mirkes:1992hu,Mirkes:1994dp},
we determine new NLO predictions. For this purpose, we use the publicly available codes \fewz (version 3.1)~\cite{Li:2012wna} 
and \dynnlo~\cite{Catani:2009sm}. These allow us to compute the full Drell-Yan cross section at 
next-to-next-to-leading (NNLO) order of QCD, when $q\bar{q}\to V$ is
the LO process. As discussed above, the contributions to the angular coefficients that we are interested in 
are at nonvanishing $q_T$, so that the order $\alpha_s^2$ in this case is only NLO. Since all
${\cal O}(\alpha_s^2)$ contributions are included in the \fewz and \dynnlo codes, we can 
therefore use these codes to extract the angular coefficients $\lambda$, $\mu$, $\nu$ at NLO,
providing a new and entirely independent calculation.

To our knowledge, such a comprehensive analysis has never been performed in the past.
Our study was very much inspired by the recent work~\cite{Peng:2015spa}, in which the
LHC results for the angular coefficients were analyzed on general theoretical grounds, 
attributing the observed violation of the Lam-Tung relation to a ``noncoplanarity'' of
the axis of the incoming partons with respect to the hadron plane, which may
be constrained by the combined Tevatron and LHC data. As the authors of~\cite{Peng:2015spa}
pointed out, the most likely physical explanation for the LHC result on the violation of the 
Lam-Tung relation is QCD radiative effects at NLO (or beyond). We indeed confirm this in our study.

We push the purely perturbative framework also to the fixed-target regime, where there have been
hardly any phenomenological analyses of the Drell-Yan angular coefficients in the context of
hard-scattering QCD. Reference~\cite{Brandenburg:1993cj} presents results at the
energy of the NA10 experiment; however the kinematics relevant at NA10 was not
properly implemented.
Of course, in the fixed-target regime $q_T$ can become quite small, smaller than, say, 
\SI{1}{GeV} or so. For such low values one does not expect fixed-order perturbation theory
to provide reliable results for cross sections, even if $Q$ is relatively large. Intrinsic 
transverse momenta of the initial partons may become relevant, among them precisely
the Boer-Mulders functions mentioned earlier. The possible role of higher-twist contributions
has been discussed as well~\cite{Brandenburg:1994wf,Zhou:2009rp}. Furthermore, as is well known, large logarithmic 
perturbative corrections of the form $\alpha_s^k \log^m(Q^2/q_T^2)/q_T^2\;$ ($m=1,\ldots,
2k-1$) appear in calculations at fixed perturbative order $k$, as a result of soft-gluon
emission. In order to describe the cross sections, one needs to resum these corrections 
to all orders in the strong coupling and also implement nonperturbative contributions
(see especially~\cite{Konychev:2005iy}, and references therein). 
As was discussed in Refs.~\cite{Boer:2006eq,Berger:2007si}, such corrections will likely 
cancel to a significant degree in the angular coefficients $\lambda$ and $\nu$, since the same 
type of leading logarithms occur in the numerator and denominator for both quantities. 
Also, it is expected~\cite{Berger:2007si} that the Lam-Tung relation will remain essentially
untouched by the soft-gluon effects. 

Thus, although clearly collinear perturbation theory at fixed-order (NLO) that we will use here
cannot provide a completely adequate framework for describing cross sections in all
kinematic regimes of interest for the angular coefficients, our results to be presented
below yield important benchmarks, in our view. In the light of the observations
concerning the soft-gluon effects mentioned above, it appears likely that fixed-order
perturbation theory will work much better for ratios of cross sections than for
the cross sections themselves. In fact, we will find that we can describe all
data sets quite well, and that we do not find any clear-cut evidence for
nontrivial additional contributions to be attributed to parton intrinsic momenta.
We stress again that QCD radiative effects are typically not considered at all 
when for example Boer-Mulders functions are extracted from data for $\nu$ (although
the conceptual framework for such a combined analysis is available~\cite{Bacchetta:2008xw}).
At the very least, our results establish the relevance of the radiative effects for
phenomenological studies of the Drell-Yan angular dependences.

Our paper is organized as follows. In Sec.~\ref{sec2} we explain how we
extract the angular coefficients from the available Drell-Yan NNLO codes. 
Section~\ref{sec3} shows our phenomenological results, and in Sec.~\ref{sec4}
we conclude our work.

\section{Extraction of angular coefficients at NLO \label{sec2}}

It is actually relatively straightforward to use the \fewz~\cite{Li:2012wna} and 
\dynnlo~\cite{Catani:2009sm} codes to determine the angular coefficients 
$\lambda,\mu,\nu$. The programs allow us to compute cross sections over suitable
ranges of any kinematic variable, providing full control over the four-momenta
of the produced particles. As already pointed out in \cite{Lam:1978pu}, the 
structure functions $W_T$, $W_L$, $W_\Delta$, $W_{\Delta\Delta}$ may be projected
out by computing the following combinations of cross sections:
\begin{align} \label{eqn: seperate Ws}
2W_{T}+W_{L} &= {\cal N}\, \frac{d\sigma}{d^{4}q} \,, \nn \\[2mm]
W_{T}-W_{L} &= \frac{8}{3}\,{\cal N}\, \brb{\frac{d\sigma}{d^{4}q}
\braa{\abs{\cos\theta}>\frac{1}{2}} - \frac{d\sigma}{d^{4}q}\braa{\abs{\cos\theta}<\frac{1}{2}}}\,,\nn \\[2mm]
W_{\Delta} &= \frac{\pi}{2}\,{\cal N}\,  \brb{\frac{d\sigma}{d^{4}q}\braa{\sin 2\theta\cos\phi>0} - 
\frac{d\sigma}{d^{4}q}\braa{\sin 2\theta\cos\phi<0}} \nn \,,\\[2mm]
W_{\Delta\Delta} &= \frac{\pi}{2}\,{\cal N}\, \brb{\frac{d\sigma}{d^{4}q}\braa{\cos 2\phi>0} - 
\frac{d\sigma}{d^{4}q}\braa{\cos 2\phi<0}} \,,
\end{align}
where ${\cal N}=12 \pi^3 (Qs/\alpha)^2$. Using Eqs.~(\ref{lmnWrel}), the angular coefficients follow immediately
from these expressions. We note that Eqs.~(\ref{eqn: seperate Ws}) are valid both for exchanged photons
and $Z$ bosons. As mentioned earlier, in cases where $Z$ bosons contribute the cross section has additional 
angular pieces; however these do not survive the integrations in Eqs.~(\ref{eqn: seperate Ws}). 

The remaining task is to determine the kinematical variables that appear in Eqs.~(\ref{eqn: seperate Ws}) from
the momenta of the outgoing leptons given in the Monte Carlo integration codes of~\cite{Li:2012wna,Catani:2009sm}.
To this end, we use that the momentum of one lepton, written in the Collins-Soper frame
as $\ell^\mu_{\mathrm{CS}}=\frac{Q}{2}(1,\sin\theta\cos\phi,
\sin\theta\sin\phi,\cos\theta)$, becomes in the hadronic c.m.s.~\cite{Arnold:2008kf} 
\begin{align}
\notag \ell_{\mathrm{cm}}^\mu		&= \frac{1}{2}
		\begin{pmatrix}
			q_{0}\braa{1+\sin\alpha \sin\theta\cos\phi} + q_L \cos\alpha \cos\theta \\[2mm]
			q_{T}\cos\varphi + Q \frac{\sin\theta}{\cos\alpha} ( \cos\phi\cos\varphi-\cos\alpha\sin\phi\sin\varphi) \\[2mm]
			q_{T}\sin\varphi + Q  \frac{\sin\theta}{\cos\alpha}  ( \cos\phi\sin\varphi+\cos\alpha\sin\phi\cos\varphi) \\[2mm]			
			q_L\braa{1+\sin\alpha \sin\theta\cos\phi} + q_0 \cos\alpha \cos\theta 
		\end{pmatrix}\,,
\end{align}
where 
\beq
\sin\alpha\equiv\frac{q_T/Q}{\sqrt{1+(q_T/Q)^2}}\,,\qquad
\cos\alpha\equiv\frac{1}{\sqrt{1+(q_T/Q)^2}}\,,
\eeq
and where $q_0$ and $q_L$ are the energy and the longitudinal component (with respect to the collision axis) 
of the virtual boson in the hadronic c.m.s., so that $q^\mu_{\mathrm{cm}}=(q_0,q_T \cos\varphi,q_T\sin\varphi,q_L)$.
To project out the combinations of trigonometric functions we need, we introduce 
\begin{align}
		\mathcal{P}_{1}^\mu\equiv
		\begin{pmatrix}
			q_L \\[1mm] 0 \\[1mm]  0 \\[1mm]  q_{0}
		\end{pmatrix}, \qquad
		\mathcal{P}_{2}^\mu \equiv q_{T}
		\begin{pmatrix}
		0 \\ \cos\varphi \\ \sin\varphi \\ 0
		\end{pmatrix} ,
		\qquad \mathcal{P}_{3}^\mu \equiv q_{T}
		\begin{pmatrix}
		0 \\ \sin\varphi \\ -\cos\varphi \\ 0
		\end{pmatrix}\,.
\end{align}
We then have
\begin{align}
\cos\theta &= - \frac{2 \,\ell_{\mathrm{cm}}\cdot \mathcal{P}_1}{(Q^2+q_T^2)\cos\alpha}\,, \nn\\[2mm]
\sin2\theta\cos\phi &= \frac{4\,\ell_{\mathrm{cm}}\cdot \mathcal{P}_1}{Q^{2}+q_{T}^{2}} 
\brb{\frac{q_{T}}{Q} + \frac{2\,\ell_{\mathrm{cm}}\cdot \mathcal{P}_2}{q_{T}Q}}\,, \nn\\[2mm]
\cos2\phi &= 1 - \frac{2\braa{\ell_{\mathrm{cm}}\cdot \mathcal{P}_3}^{2}}{q_{T}^{2}
\brb{\frac{Q^2}{4}-\frac{\braa{\ell_{\mathrm{cm}}\cdot \mathcal{P}_1}^{2}}{Q^{2}+q_{T}^{2}}}}\,.\label{ellP}
\end{align}
The four-momentum of the lepton in the hadronic c.m.s. is provided in the Monte Carlo integration codes,
while that of the virtual boson is fixed by the external kinematics.
Writing $\ell^\mu_{\mathrm{cm}}=(\ell_{\mathrm{cm}}^0,\ell_{\mathrm{cm}}^1,\ell_{\mathrm{cm}}^2,\ell_{\mathrm{cm}}^3)$,
we have 
\begin{align}
\ell_{\mathrm{cm}}\cdot \mathcal{P}_1 &= q_L \,\ell_{\mathrm{cm}}^0-q_0\, \ell_{\mathrm{cm}}^3 \,,\nn \\[2mm]
\ell_{\mathrm{cm}}\cdot \mathcal{P}_2 &= -q_{T}\braa{\ell_{\mathrm{cm}}^1\,\cos\varphi + \ell_{\mathrm{cm}}^2\,\sin\varphi} \,,\nn\\[2mm]
\ell_{\mathrm{cm}}\cdot \mathcal{P}_3 &= q_{T}\braa{\ell_{\mathrm{cm}}^2\,\cos\varphi - \ell_{\mathrm{cm}}^1\,\sin\varphi}\,.
\end{align}
Inserting these expressions into Eqs.~(\ref{ellP}), one can now easily implement the appropriate cuts
in the codes so that the structure functions $W_T$, $W_L$, $W_\Delta$, $W_{\Delta\Delta}$ can be extracted via 
Eqs.~(\ref{eqn: seperate Ws}). 

\section{Comparison to data \label{sec3}}

We now present comparisons of the theoretical predictions at LO and NLO to the available experimental 
data for the angular coefficients $\lambda$ and $\nu$. We do not show any results for the coefficient $\mu$
which comes out always extremely small and in fact usually consistent with zero both in the theoretical calculation
and in experiment, within the respective uncertainties. We first note that we have validated our technique for 
extracting the Drell-Yan angular
coefficients from the \fewz (version 3.1)~\cite{Li:2012wna} and \dynnlo~\cite{Catani:2009sm} codes by writing a completely 
independent LO code. We have found perfect agreement between this code and the LO results we extracted from 
\fewz and \dynnlo. In the figures below, the LO curves will always refer to those from our own code. 
We also note that the NLO results we show in the following have all been obtained with the \fewz code. 
We have compared to the results of \dynnlo and found excellent consistency of the two codes both 
at LO and NLO. 

\begin{figure}[t!]
	\centering
	\includegraphics[width=0.9\textwidth]{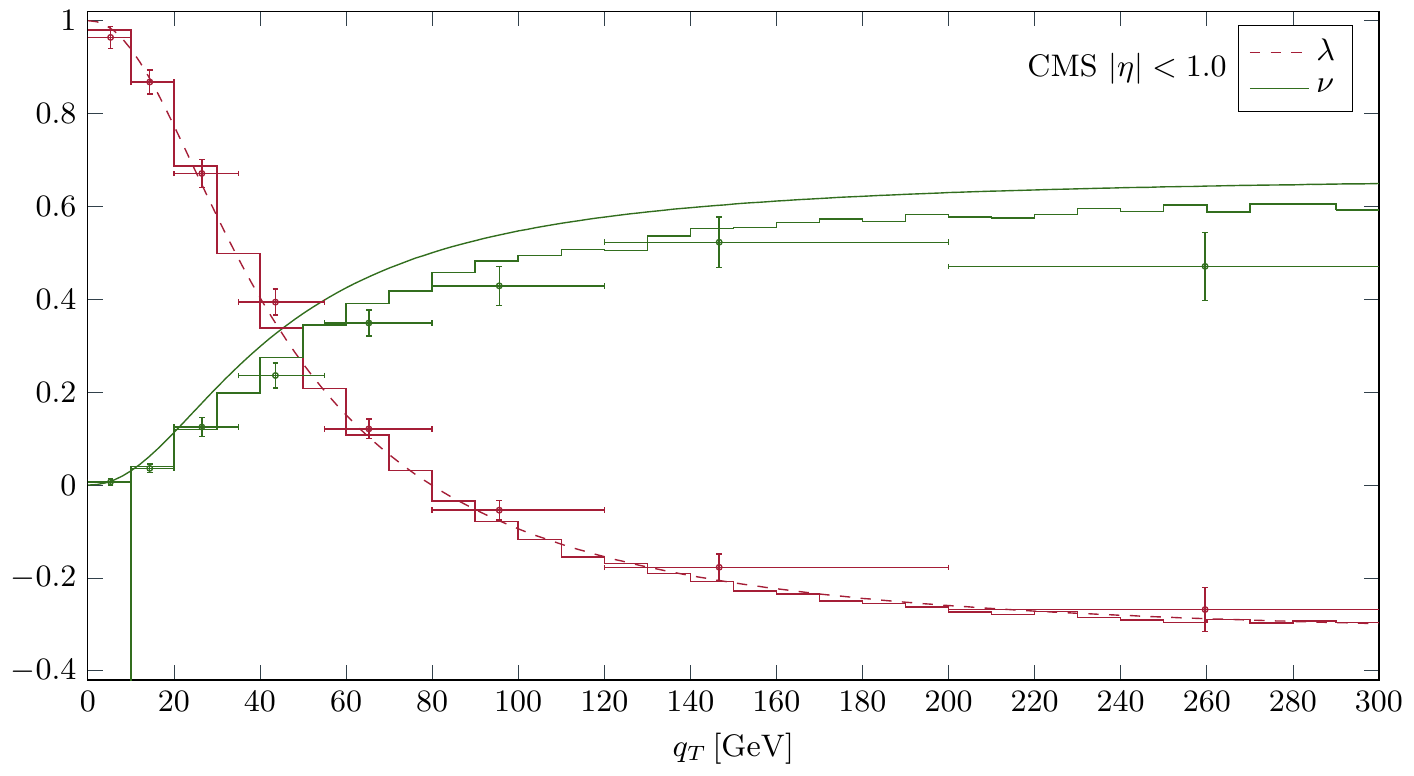}
	\vspace*{0.cm}
	\caption{{\sl Comparison of LO (lines) and NLO (\fewz~\cite{Li:2012wna}, histograms) theoretical results  
	to the CMS data~\cite{Khachatryan:2015paa} for the angular coefficients $\lambda$ and $\nu$ 
	taken at $\sqrt{s}=\SI{8}{TeV}$. We have integrated over $\num{81}\leq Q\leq \SI{101}{GeV}$ and over a midrapidity interval $|\eta|<1$ of the 
	virtual boson.}}
	\label{fig1}
\vspace*{14mm}
	\centering
	\includegraphics[width=0.9\textwidth]{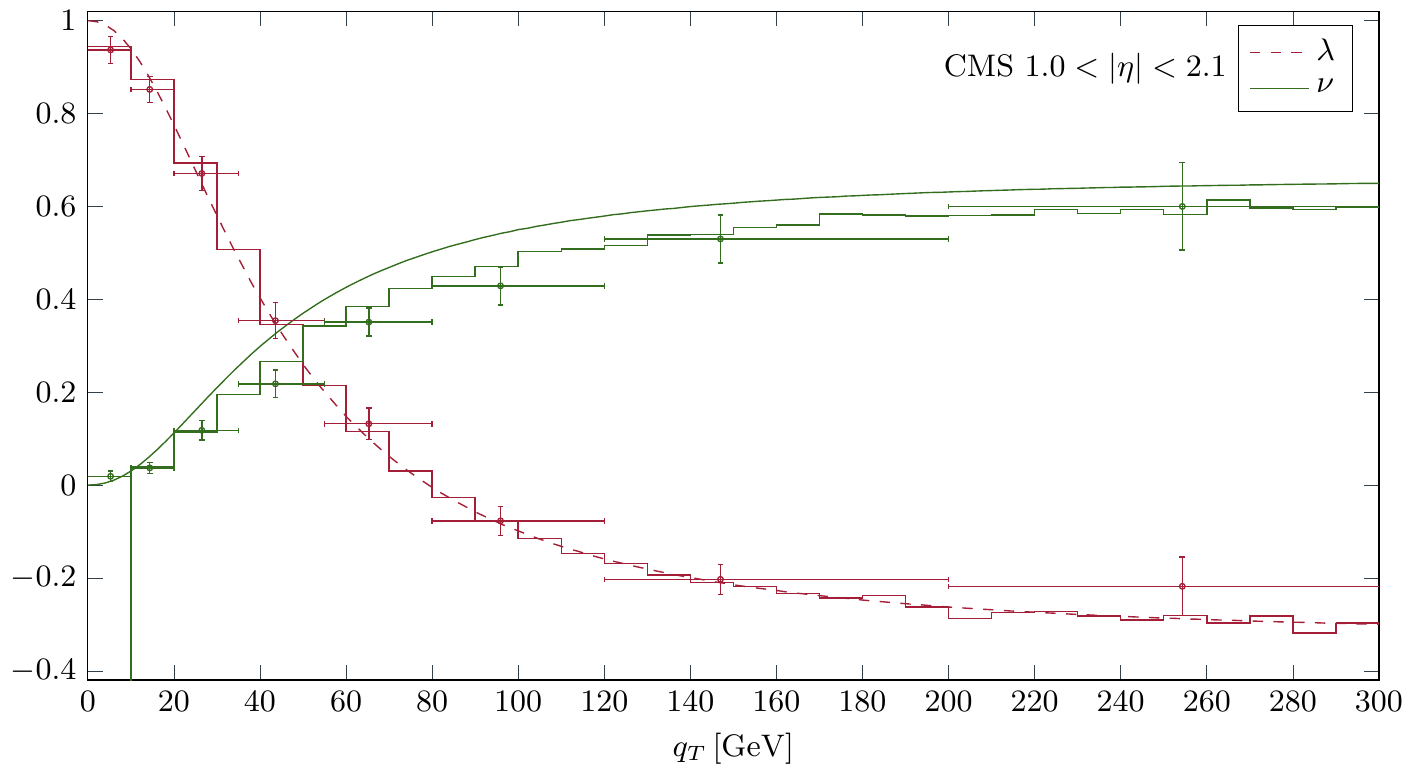}
	\vspace*{-0.cm}
	\caption{{\sl Same as Fig.~\ref{fig1}, but for a more forward/backward rapidity interval $1<|\eta|< 2.1$.}}
	\label{fig2}
\end{figure}

Although the implementation of Eqs.~(\ref{eqn: seperate Ws}) and the relevant kinematics into
the \fewz or \dynnlo codes is relatively straightforward, the computational load for performing
a comprehensive comparison of the data with NLO theory is very large. To obtain the NLO results
presented below, we have run an equivalent of one \SI{3.20}{GHz} Intel Quad-Core i5-3470 CPU using 
all of its cores for about 2 years. In order to collect sufficiently high statistics at very high values of $q_T$, where the cross section drops very rapidly, we have performed dedicated runs for which we have implemented cuts on the low-$q_{T}$ region, forcing the Monte Carlo integration to sample high $q_{T}$. We also note that typically the result for the lowest-$q_T$ bin
is unreliable, since this bin contains the (NNLO) contributions at $q_T=0$. Nonetheless, our results are 
sufficiently accurate in all regions of interest and thus allow us to derive solid conclusions. 
We mention that we also had to modify the codes to accommodate pion beams and nuclear (deuteron/tungsten)
targets. This implementation was always checked against our own LO code.

\begin{figure}[t!]
	\centering
	\includegraphics[width=0.9\textwidth]{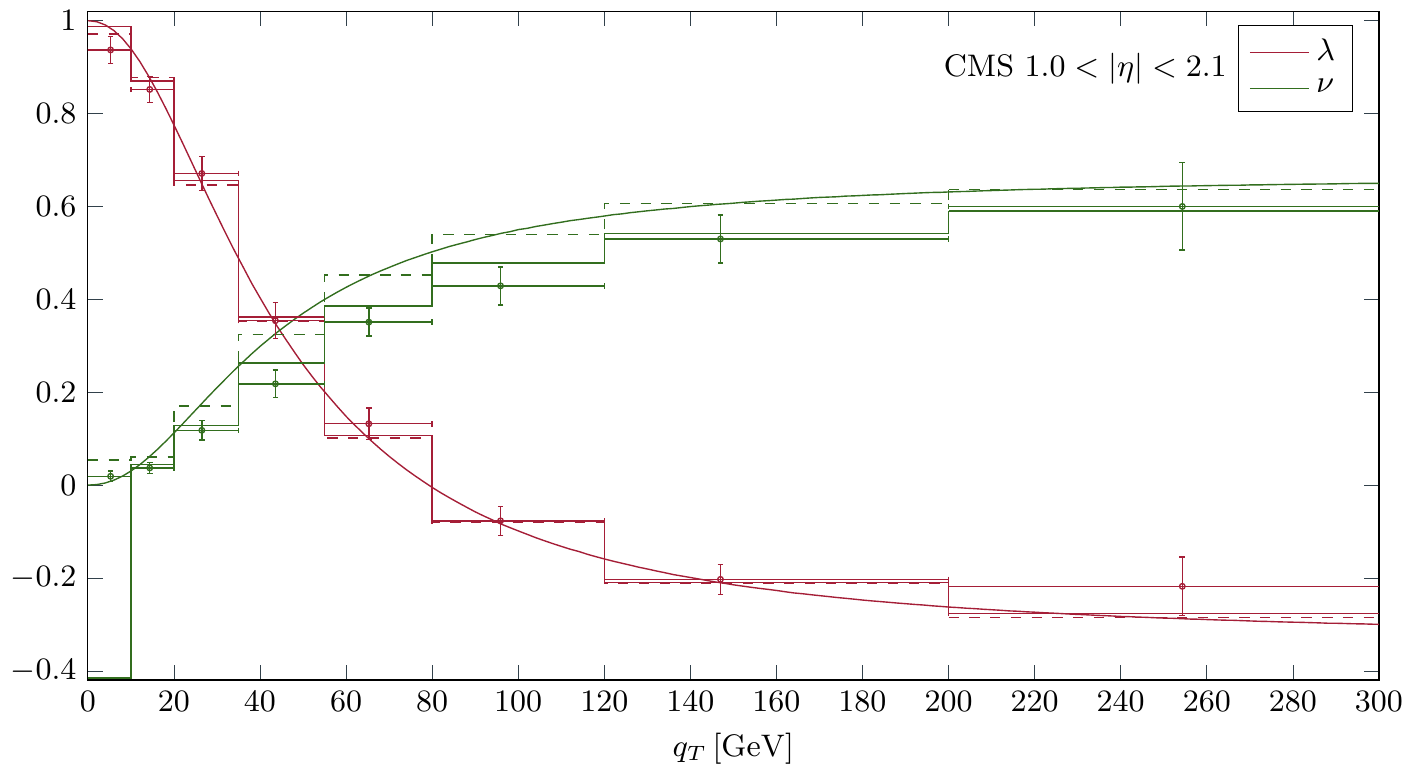}
	\vspace*{0.cm}
	\caption{{\sl Same as Fig.~\ref{fig2}, but with the NLO theoretical results integrated 
	over the eight $q_T$ bins used by CMS. In this figure, the dashed histograms show the LO results and the 
	solid ones the NLO results. To guide the eye, we also show the LO results from Fig.~\ref{fig2} as smooth lines.}}
	\label{fig1X}
\vspace*{14mm}
	\centering
	\includegraphics[width=0.9\textwidth]{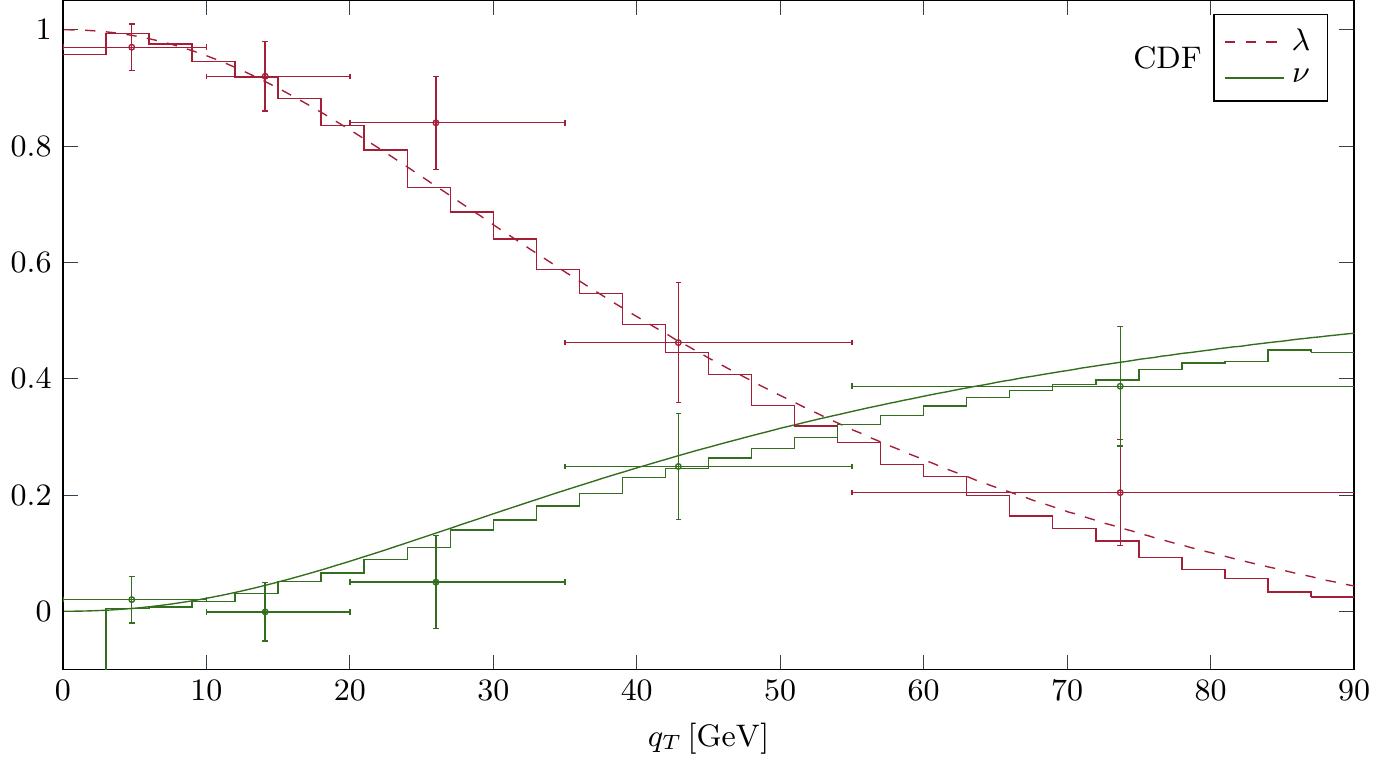}
	\vspace*{-0.cm}
	\caption{{\sl Comparison of LO (lines) and NLO (\fewz~\cite{Li:2012wna}, histograms) theoretical results 
	to the CDF data~\cite{Aaltonen:2011nr} for the angular coefficients $\lambda$ and $\nu$ 
	taken in $p\bar{p}$ scattering at $\sqrt{s}=\SI{1960}{GeV}$. We have integrated over $\num{66}\leq Q\leq \SI{116}{GeV}$	and over $|\eta|<3.6$ of the virtual boson.}}
	\label{fig3}
	\vspace*{-8mm}
\end{figure}

Throughout this paper, we use the parton distribution functions of the proton 
of Ref.~\cite{Martin:2009iq}, adopting their NLO (LO) set for the NLO (LO) calculation. The choice
of parton distributions has a very small effect on the Drell-Yan angular coefficients. When dealing
with nuclear targets (tungsten was used for all of the pion scattering experiments and deuterons for one set of 
E866 measurements) 
we compute the parton distributions of the nucleus just by considering the relevant isospin relations for protons and
neutrons, averaging over the appropriate proton and neutron number. We do not add any other nuclear 
effects. For the parton distributions of the pion, we use the set in~\cite{Sutton:1991ay}; the set in~\cite{Gluck:1991ey}
would give very similar results. Finally, our choice for the factorization and renormalization scales will always be $\mu=Q$. 
We have checked that other possible scale choices such as $\mu=\sqrt{Q^2+q_T^2}$ do not change
the results for the angular coefficients significantly even at LO, making an impact of at most a few percent, and only
at high values of $q_T$. Here we have simultaneously varied the scales in the cross sections appearing in 
the numerators {\it and} in the denominators of the angular coefficients; relaxing this condition one would likely
be able to generate a larger dependence on the choice of scale. On the other hand, as is known from
previous calculations~\cite{Li:2012wna,Catani:2009sm}, the scale dependence of the Drell-Yan cross section
is overall much reduced at higher orders anyway.

We present our results essentially in the order of decreasing energy, starting with a comparison to the high-energy
collider data from the LHC~\cite{Khachatryan:2015paa} and Tevatron~\cite{Aaltonen:2011nr}.
The reason is that for these data sets $Q$ is very large, $Q\sim m_Z$, so that perturbative methods should be
well justified. The transverse momentum $q_T$ varies over a broad range, taking low values as well as values of 
order $Q$. At the lower end, where $q_T\ll Q$, it may well be necessary to perform an all-order resummation of
perturbative double logarithms in $q_T/Q$ in order to describe the Drell-Yan cross section properly.
However, as mentioned in the Introduction, such logarithms are expected to cancel to a large extent
in the angular coefficients~\cite{Boer:2006eq,Berger:2007si}. Thus, if ever fixed-order perturbative QCD 
predictions are able to provide an adequate description of the angular coefficients, it should be 
in the kinematic regimes explored at the LHC and Tevatron.

Figures~\ref{fig1} and~\ref{fig2} show our results for $\lambda$ and $\nu$ compared to the CMS data~\cite{Khachatryan:2015paa},
for two separate bins in the rapidity of the virtual boson,
\beq
\eta\,\equiv\,\frac{1}{2}\log\frac{q_0+q_L}{q_0-q_L}\,.
\eeq
We note that CMS presents their data in terms of
a different set of angular coefficients termed $A_0$, $A_1$, $A_2$, $A_3$, which are directly related to the coefficients
we use here. In particular, we have $\lambda=(2-3A_0)/(2+A_0)$ and $\nu=2A_2/(2+A_0)$. As in Ref.~\cite{Peng:2015spa},
in order to present a full comparison in terms of $\lambda$ and $\nu$, we transform the experimental data correspondingly. 
Here we have propagated the experimental uncertainties, albeit without taking into account any 
correlations. The lines in the figures show our LO results for the coefficients. As one can see, they qualitatively
follow the trend of the data, but for the coefficient $\nu$ a clear deviation between data and LO theory
is observed. This is precisely the finding also emphasized in Ref.~\cite{Peng:2015spa} where it was argued
(without explicit NLO calculation)
that the discrepancy ought to be related to higher-order QCD effects. Indeed, this is what we find. The NLO
results (histograms) show a markedly better agreement with the data, which in fact is nearly perfect.
The coefficient $\lambda$, on the other hand, changes only marginally from LO to NLO. As is visible in the figures, the results at very high values of $q_T$ are numerically less accurate, as shown by the 
somewhat erratic behavior of the histograms. In order to collect higher statistics, we have also performed runs for which
we integrated over only eight $q_T$ bins, choosing exactly the ones used in the experimental analysis. 
The corresponding results are shown in Fig.~\ref{fig1X} for the range $1<|\eta|< 2.1$. Our goal was to 
make sure that the numerical uncertainty for these bins is much smaller than the experimental one {\it even} in the bin 
at highest $q_T$. The figure once more impressively shows how NLO theory leads to an excellent description
of the CMS data. 

It is interesting 
to note that NLO \fewz results were also shown in the CMS paper~\cite{Khachatryan:2015paa}. However,
the agreement with the data for the coefficient $A_2$ (which multiplies the $\cos 2\phi$ 
dependence of the cross section) reported there appears to be not quite as good as the one
we find for our coefficient $\nu$. It is conceivable that our computation of the coefficients via Eqs.~\eqref{eqn: seperate Ws}
is numerically more stable.

We next turn to the comparison to the CDF data~\cite{Aaltonen:2011nr} taken in $p\bar{p}$ collisions at 
$\sqrt{s}=\SI{1960}{GeV}$ at the Tevatron. The results are shown in Fig.~\ref{fig3}. We observe that both 
the LO and the NLO results are in good agreement with the data, NLO doing a bit better overall. 
Both coefficients $\lambda$ and $\nu$ decrease slightly when going to NLO. For $\nu$, this effect is 
less pronounced than for the LHC case, which may be attributed to a much stronger contribution
by the $q\bar{q}$ channel in the present $p\bar{p}$ case, which receives smaller radiative corrections. 
Again, this feature was predicted phenomenologically in Ref.~\cite{Peng:2015spa}.

\begin{figure}[t!]
	\vspace*{-0.cm}
	\centering
	\includegraphics[width=0.9\textwidth]{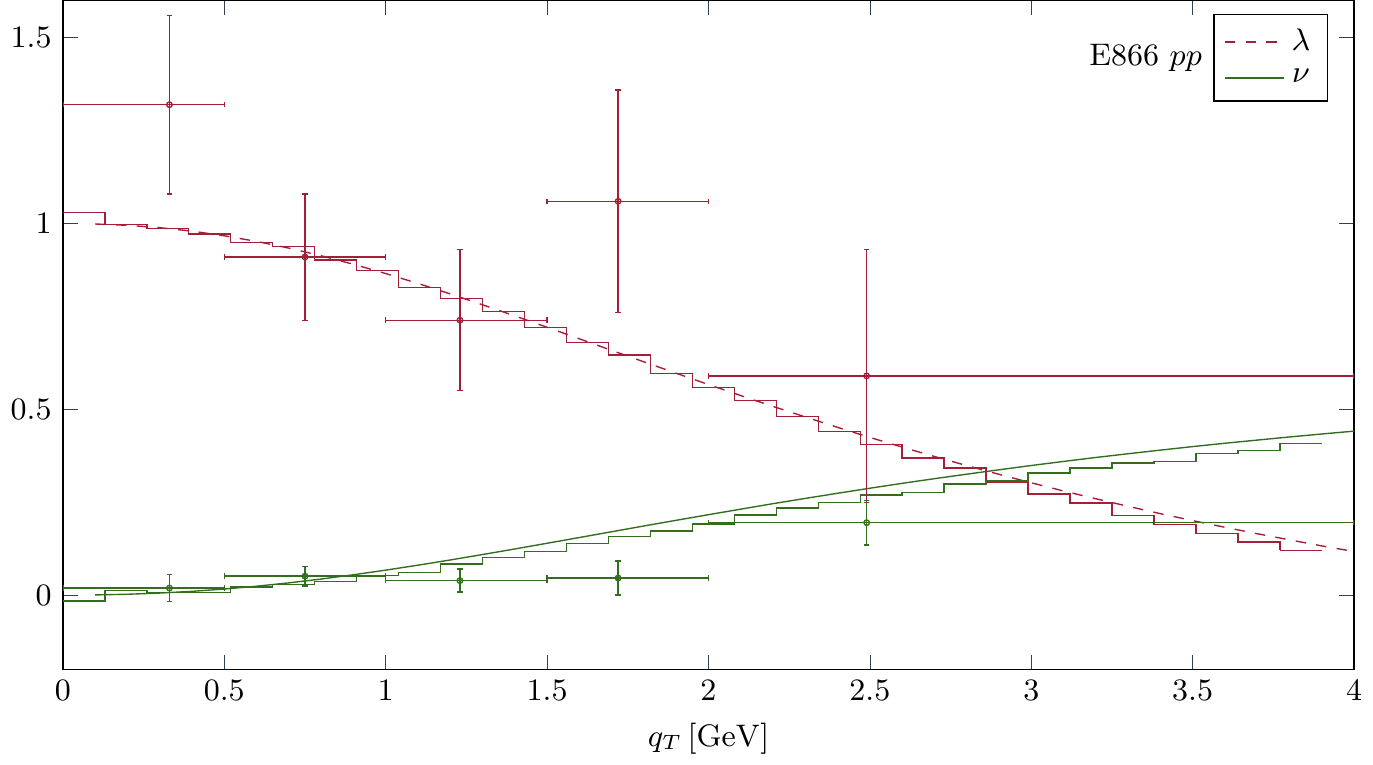}
	\vspace*{-0.cm}
	\caption{{\sl Comparison of LO (lines) and NLO (\fewz~\cite{Li:2012wna}, histograms) theoretical results 
	 to the $pp$ scattering data from E866~\cite{Zhu:2008sj} taken
	with an \SI{800}{GeV} beam. Error bars are statistical only. 
	We have integrated over the mass range $\num{4.5}\leq Q\leq \SI{15}{GeV}$, excluding 
the bottomonium region $\num{9}\leq Q\leq \SI{10.7}{GeV}$. We have also integrated over $0\leq x_F\leq \num{0.8}$, where
$x_F=2q_L/\sqrt{s}$ is the Feynman variable and is counted as positive in the forward direction of the beam.}}
	\label{fig4}
\vspace*{10mm}
	\centering
	\includegraphics[width=0.9\textwidth]{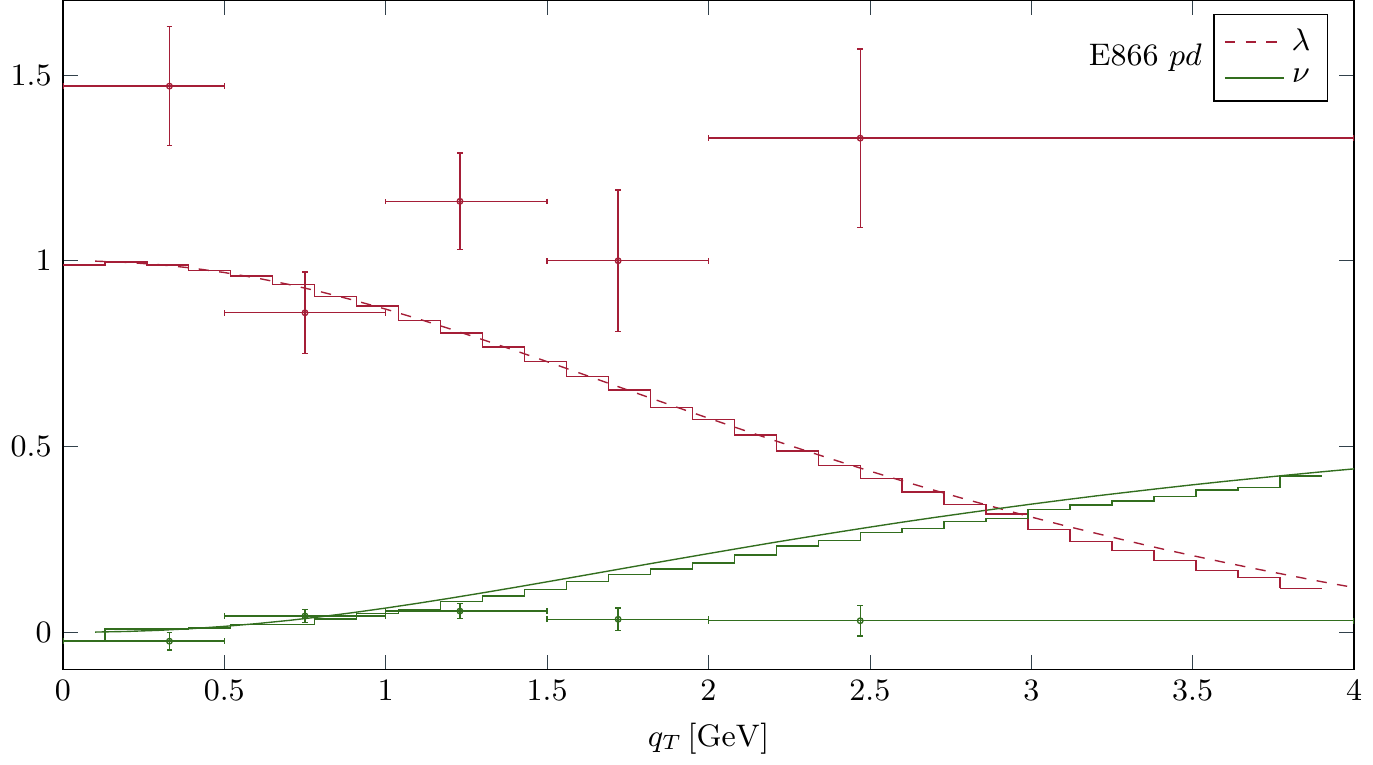}
	\vspace*{-0.cm}
	\caption{{\sl Same as Fig.~\ref{fig4}, but for $pd$ scattering. Data are from Ref.~\cite{Zhu:2006gx}}}
	\label{fig5}
	\vspace*{-4mm}
\end{figure}

We now consider the fixed-target regime, where we start with a comparison to the 
Fermilab E866/NuSea data taken with an \SI{800}{GeV} proton beam in $pp$~\cite{Zhu:2008sj}
and $pd$~\cite{Zhu:2006gx} scattering. The comparisons to the two data sets are shown
in Figs.~\ref{fig4} and~\ref{fig5}. We first note that the $pp$ data are overall in much 
better agreement with the theoretical curves than the $pd$ ones. For $pp$ scattering,
the coefficient $\lambda$ is well described,
given the relatively large experimental uncertainties. There is a slight trend in the data for the 
coefficient $\nu$ to be lower than the theoretical prediction. The NLO corrections in fact 
provide a slight improvement here. For $pd$ scattering, the two data points for $\nu$ at the 
highest $q_T$ are clearly below theory even at NLO. The coefficient $\lambda$ is not well 
described, neither at LO nor at NLO. An important point to note in this context is the 
positivity constraint~\cite{Lam:1978pu} 
\beq
W_L\geq 0\,,
\eeq
which immediately implies
\beq \label{constraint}
\lambda \leq 1 \,.
\eeq 
This condition is completely general and relies only on the hermiticity of the neutral
current. It is interesting to observe that the $pd$ data shown in Fig.~\ref{fig4} are only in
borderline agreement with this positivity constraint. 

\begin{figure}
	\centering
	\includegraphics[width=0.9\textwidth]{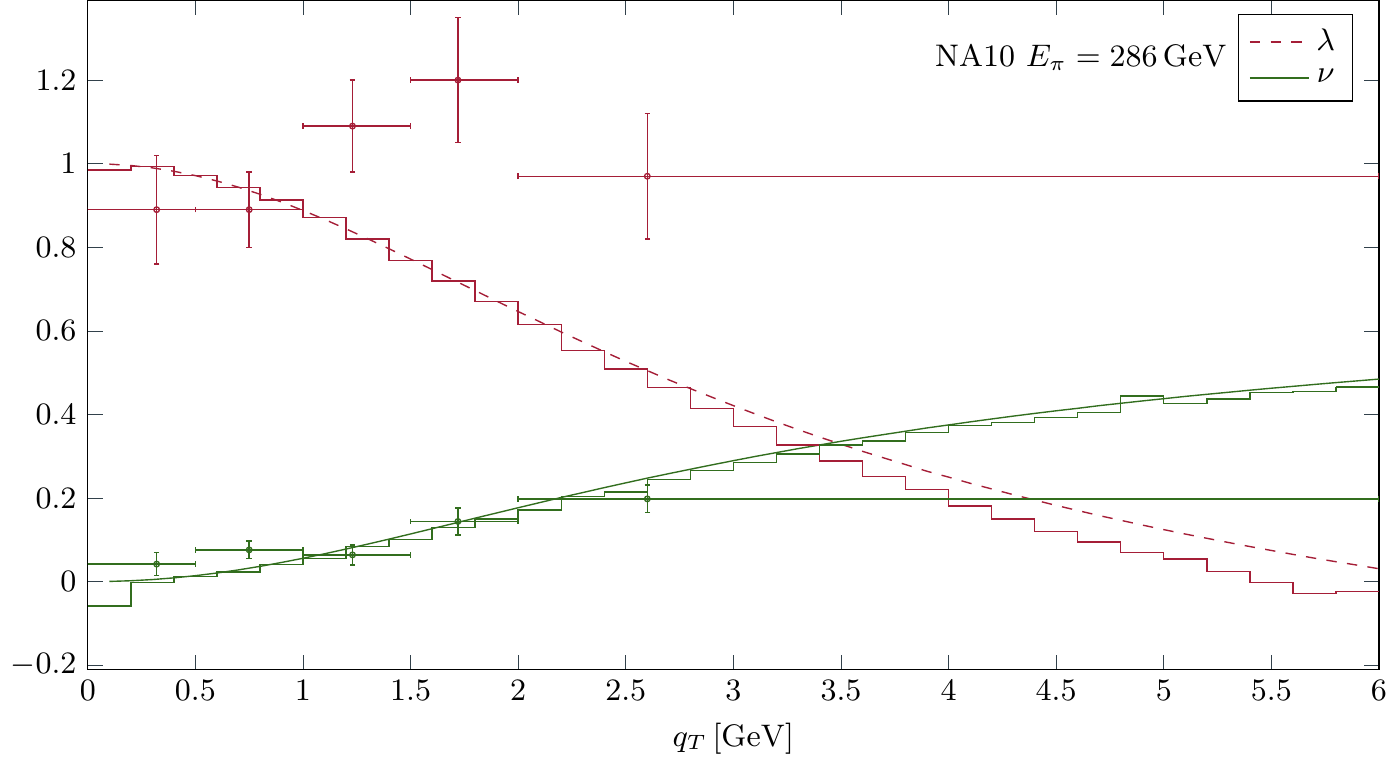}
	\vspace*{-0.cm}
	\caption{{\sl Comparison of LO (lines) and NLO (\fewz~\cite{Li:2012wna}, histograms) theoretical results 
	 to the $\pi+$tungsten scattering data from NA10~\cite{Guanziroli:1987rp} taken
	with pion beam energy $E_\pi=\SI{286}{GeV}$. Error bars are statistical only. We have integrated over the mass range $Q\geq \SI{4}{GeV}$, excluding 
the bottomonium region $\num{8.5}\leq Q\leq \SI{11}{GeV}$. We have also implemented the cut $0\leq x_\pi\leq \num{0.7}$, 
where $x_\pi=\frac{1}{2}(x_F+\sqrt{x_F^2+4Q^2/s})$ with $x_F=2q_L/\sqrt{s}$ the Feynman variable, which 
is counted as positive in the forward direction of the pion beam.}}
	\label{fig6}
\vspace*{8mm}	
 \centering
	\includegraphics[width=0.9\textwidth]{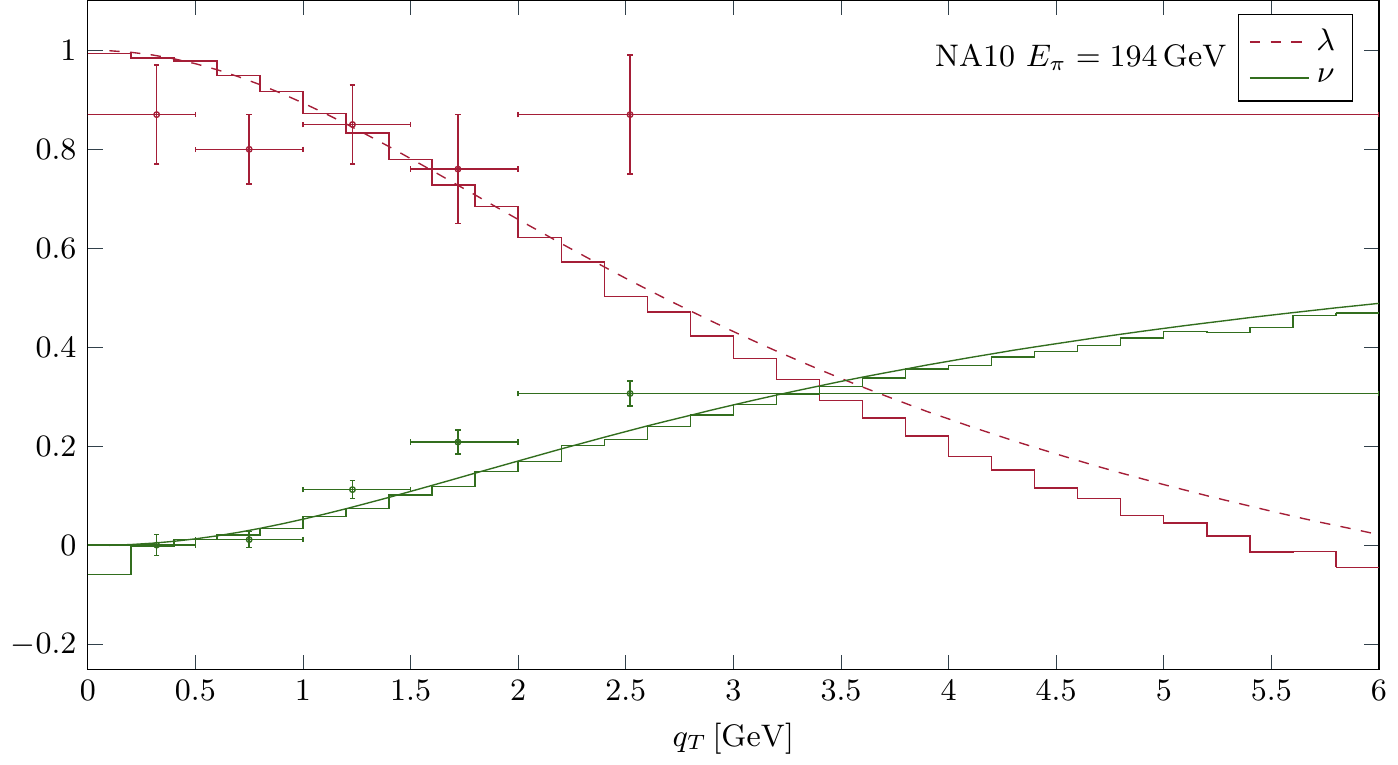}
	\vspace*{-0.cm}
	\caption{{\sl Same as Fig.~\ref{fig6}, but at pion energy $E_\pi=\SI{194}{GeV}$ and integrated over $Q\geq \SI{4.05}{GeV}$.}}
	\label{fig7}
	\vspace*{-4mm}
\end{figure}

\begin{figure}
	\centering
	\includegraphics[width=0.9\textwidth]{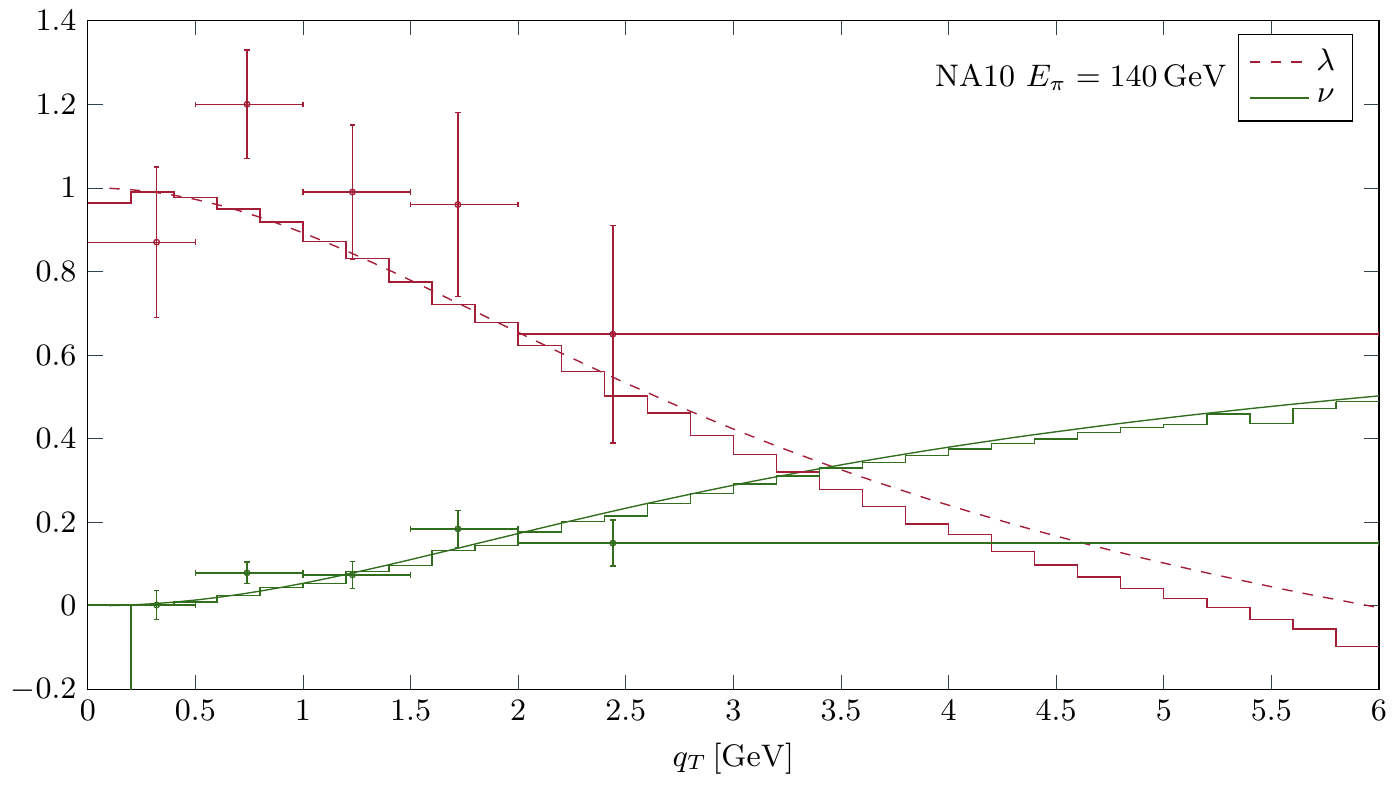}
	\vspace*{-0.cm}
	\caption{{\sl Same as Fig.~\ref{fig6}, but at pion energy $E_\pi=\SI{140}{GeV}$.}}
	\label{fig8}
\vspace*{8mm}	
\centering
	\includegraphics[width=0.9\textwidth]{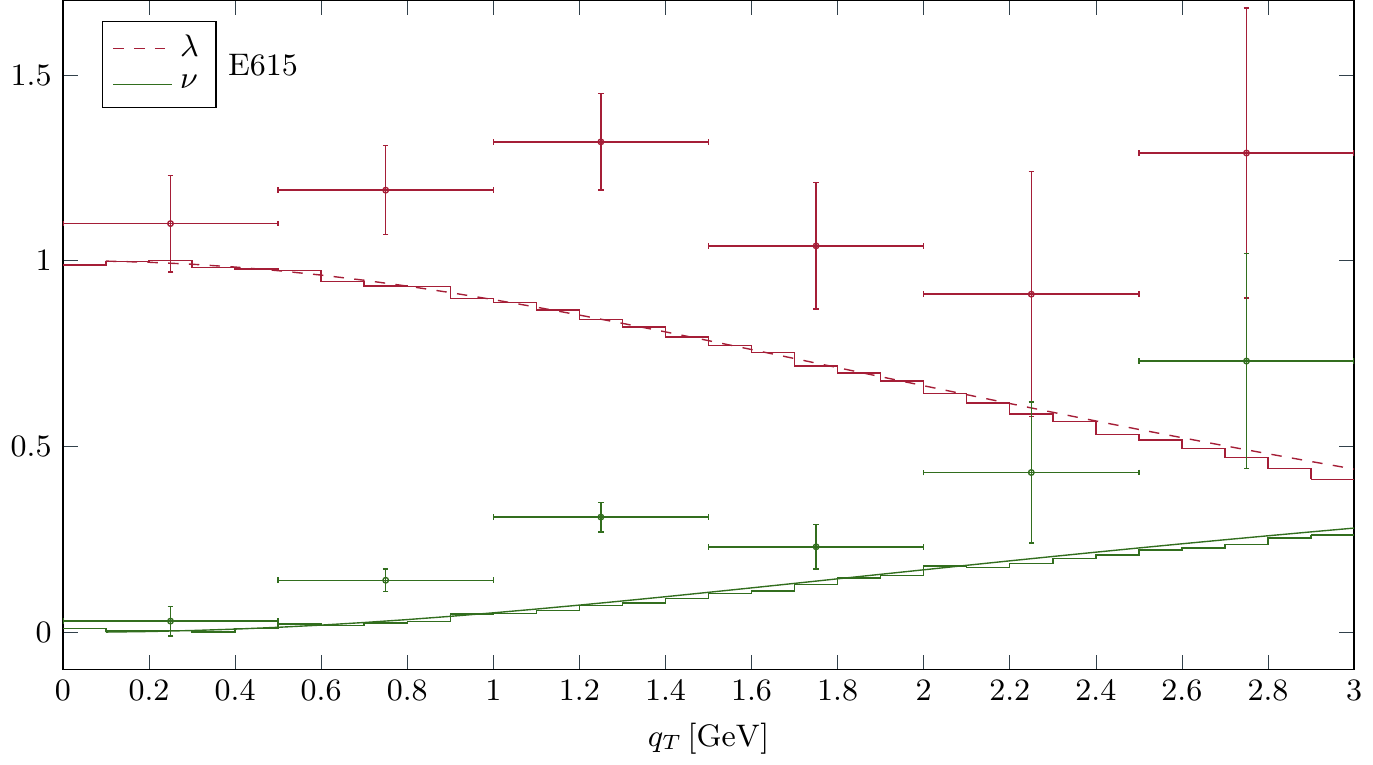}
	\vspace*{-0.cm}
	\caption{{\sl Comparison of LO (lines) and NLO (\fewz~\cite{Li:2012wna}, histograms) theoretical results 
	 to the $\pi+$tungsten scattering data from E615~\cite{Conway:1989fs} taken
	with pion beam energy $E_\pi=\SI{252}{GeV}$. We have integrated over the mass range $\num{4.05}\leq Q\leq \SI{8.55}{GeV}$. 
	We have also implemented the cuts $0\leq x_F\leq 1$ and $\num{0.2}\leq x_\pi\leq 1$, 
where $x_\pi=\frac{1}{2}(x_F+\sqrt{x_F^2+4Q^2/s})$ with $x_F=2q_L/\sqrt{s}$ the Feynman variable, which 
is counted as positive in the forward direction of the pion beam.}}
	\label{fig9}
	\vspace*{-4mm}
\end{figure}

Going further down in energy, we finally discuss the data from the $\pi+$tungsten scattering experiments 
NA10~\cite{Guanziroli:1987rp} and E615~\cite{Conway:1989fs}. NA10 used three different 
energies for the incident pions, $E_\pi=\SIlist{286;194;140}{GeV}$, while E615 operated a pion beam with energy \SI{252}{GeV}. 
Figures~\ref{fig6}--\ref{fig8} show the comparisons of our LO and NLO results for $\lambda$ and $\nu$ to the 
NA10 data. The NLO corrections are overall small for $\nu$, but for $\lambda$ they become more pronounced toward
larger $q_T$. We note that NLO results for one of the NA10 energies were also reported in Ref.~\cite{Brandenburg:1993cj},
where however not the appropriate kinematical regime in $Q$ was chosen, leading to an underestimate of $\nu$
which has unfortunately given rise to the general notion in the literature that perturbative QCD cannot 
describe the Drell-Yan angular coefficients. We also note that for the kinematics used 
in~\cite{Brandenburg:1993cj} the NLO corrections appear to be somewhat smaller than the ones we find here. 
The three cases shown in Figs.~\ref{fig6}--\ref{fig8} have in common that the data for $\nu$ are well described, 
perhaps slightly less so for the pion energy \SI{194}{GeV}. The experimental uncertainties for the coefficient $\lambda$
are very large, and it is not possible to draw solid conclusions from the comparison. We note that
wherever there are tensions between data and theory concerning $\lambda$, the data tend to 
lie uncomfortably close to (or even above) the positivity constraint $\lambda \leq 1$. 

In case of E615, we find the results shown in Fig.~\ref{fig9}. We observe that neither the description
of $\lambda$ nor that of $\nu$ is good. The NLO corrections are overall small and thus do not
change this picture. It is clear that on the basis of the data one would derive a significant
violation of the Lam-Tung relation~(\ref{LT}), since $\lambda$ and $\nu$ both enter the 
relation with the same sign, and the data for both $\lambda$ and $\nu$ are higher than theory 
(the latter satisfying the relation at LO). It is worth pointing out, however, that the experimental uncertainties 
are large and, more importantly, again the data show a certain tension with respect to the 
positivity limit~(\ref{constraint}). 

\section{Conclusions \label{sec4}}

We have presented detailed and exhaustive comparisons of data for the Drell-Yan lepton angular
coefficients $\lambda$ and $\nu$ to LO and NLO perturbative-QCD calculations. To obtain
NLO results, we have employed public codes that allow us to compute the full Drell-Yan cross
section at NNLO, and in which the angular pieces we are interested in are contained. 

Our numerical results show that overall perturbative QCD is able to describe the
experimental data quite well. For the recent LHC data the agreement is very good, 
when the NLO corrections are taken into account. This finding is in line with 
arguments made in the recent literature~\cite{Peng:2015spa}. Also the Tevatron
data are very well described at NLO. Toward the fixed-target regime, we again 
find an overall good agreement, with possible exceptions for the E866 $pd$ data 
set for $\nu$ at high $q_T$ and for the E615 data. We remark that the latter data
set carries large uncertainties and also hints at tensions with the positivity constraint
$\lambda\leq 1$. 

To be sure, the description of the cross sections that enter the angular coefficients
requires input beyond fixed-order QCD perturbation theory, notably in terms of 
resummations of logarithms in $q_T/Q$ and of transverse-momentum dependent
parton distributions. On the other hand, based on the angular coefficients alone, in our 
view there is no convincing evidence for any effects other than the ones we have
considered here. In particular, we argue that one should dispel the myth that 
perturbative QCD is not able to describe the Drell-Yan angular coefficients, 
which in fact has been iterated over and over in the literature. While we most
certainly do not wish to exclude the presence of contributions by the Boer-Mulders 
effect in the $\cos2\phi$ part of the angular distribution, it is also clear from our
study that future phenomenological studies of the effect should incorporate the QCD 
radiative effects.

We finally stress that our results clearly make the case for
new precision data for the Drell-Yan angular coefficients that
would allow to convincingly establish whether there are departures from the 
``plain'' QCD radiative effects we have considered here. 
We hope that such data will be forthcoming from measurements at the 
COMPASS~\cite{Chiosso:2015naa} or E906~\cite{Nakano:2016jdm} experiments,
or possibly at RHIC.

\section*{Acknowledgments} 
We thank Jen-Chieh Peng for useful discussions and for communications on the E866 data. 
We also acknowledge helpful discussions with Peter Schweitzer and Marco Stratmann.
This work was supported in part by the Deutsche Forschungsgemeinschaft (Grant No. VO 1049/1).
The authors acknowledge support by the state of Baden-W\"{u}rttemberg through bwHPC.




\begin{thebibliography}{99}

\bibitem{Collins:1977iv} 
  J.~C.~Collins and D.~E.~Soper,
  Phys.\ Rev.\ D {\bf 16}, 2219 (1977).
  
\bibitem{Lam:1978pu} 
  C.~S.~Lam and W.~K.~Tung,
  Phys.\ Rev.\ D {\bf 18}, 2447 (1978).
    
\bibitem{Boer:2006eq} 
  D.~Boer and W.~Vogelsang,
  Phys.\ Rev.\ D {\bf 74}, 014004 (2006)
  [hep-ph/0604177].
  
\bibitem{Berger:2007si} 
  E.~L.~Berger, J.~W.~Qiu and R.~A.~Rodriguez-Pedraza,
  Phys.\ Lett.\ B {\bf 656}, 74 (2007)
  [arXiv:0707.3150 [hep-ph]];
  Phys.\ Rev.\ D {\bf 76}, 074006 (2007)
  [arXiv:0708.0578 [hep-ph]].  

\bibitem{Peng:2014hta} 
 {\it For review, see} J.~C.~Peng and J.~W.~Qiu,
  Prog.\ Part.\ Nucl.\ Phys.\  {\bf 76}, 43 (2014)
  [arXiv:1401.0934 [hep-ph]].
  
\bibitem{Guanziroli:1987rp} 
  M.~Guanziroli {\it et al.} (NA10 Collaboration),
  Z.\ Phys.\ C {\bf 37}, 545 (1988).
    
\bibitem{Conway:1989fs} 
  J.~S.~Conway {\it et al.},
  Phys.\ Rev.\ D {\bf 39}, 92 (1989).
  
\bibitem{Zhu:2006gx} 
  L.~Y.~Zhu {\it et al.} (FNAL E866/NuSea Collaboration),
  Phys.\ Rev.\ Lett.\  {\bf 99}, 082301 (2007)
  [hep-ex/0609005].

\bibitem{Zhu:2008sj} 
  L.~Y.~Zhu {\it et al.} (FNAL E866/NuSea Collaboration),
  Phys.\ Rev.\ Lett.\  {\bf 102}, 182001 (2009)
  [arXiv:0811.4589 [nucl-ex]].

\bibitem{Aaltonen:2011nr} 
  T.~Aaltonen {\it et al.} (CDF Collaboration),
  Phys.\ Rev.\ Lett.\  {\bf 106}, 241801 (2011)
  [arXiv:1103.5699 [hep-ex]].
  
\bibitem{Khachatryan:2015paa} 
  V.~Khachatryan {\it et al.} (CMS Collaboration),
  Phys.\ Lett.\ B {\bf 750}, 154 (2015)
  [arXiv:1504.03512 [hep-ex]].  
 
\bibitem{Boer:1999mm} 
  D.~Boer,
  Phys.\ Rev.\ D {\bf 60}, 014012 (1999)
  [hep-ph/9902255].

\bibitem{Boer:1997nt} 
  D.~Boer and P.~J.~Mulders,
  Phys.\ Rev.\ D {\bf 57}, 5780 (1998)
  [hep-ph/9711485]. 

\bibitem{Collins:2002kn} 
 S.~J.~Brodsky, D.~S.~Hwang and I.~Schmidt,
  Phys.\ Lett.\ B {\bf 530}, 99 (2002)
  [hep-ph/0201296]; 
  Nucl.\ Phys.\ B {\bf 642}, 344 (2002)
  [hep-ph/0206259]; 
  J.~C.~Collins,
  Phys.\ Lett.\ B {\bf 536}, 43 (2002)
  [hep-ph/0204004].

\bibitem{Barone:2010gk} 
  V.~Barone, S.~Melis and A.~Prokudin,
  Phys.\ Rev.\ D {\bf 82}, 114025 (2010)
  [arXiv:1009.3423 [hep-ph]].

\bibitem{Lu:2011mz} 
  Z.~Lu and I.~Schmidt,
  Phys.\ Rev.\ D {\bf 84}, 094002 (2011)
  [arXiv:1107.4693 [hep-ph]].

\bibitem{Pasquini:2014ppa} 
D.~Boer, S.~J.~Brodsky and D.~S.~Hwang,
  Phys.\ Rev.\ D {\bf 67}, 054003 (2003)
  [hep-ph/0211110];
 Z.~Lu and B.~Q.~Ma,
  Phys.\ Lett.\ B {\bf 615}, 200 (2005)
  [hep-ph/0504184]; 
  B.~Pasquini and P.~Schweitzer,
  Phys.\ Rev.\ D {\bf 90}, 014050 (2014)
  [arXiv:1406.2056 [hep-ph]].
    
\bibitem{Lam:1978zr} 
  C.~S.~Lam and W.~K.~Tung,
  Phys.\ Lett.\ B {\bf 80}, 228 (1979).

\bibitem{Collins:1978yt} 
  J.~C.~Collins,
  Phys.\ Rev.\ Lett.\  {\bf 42}, 291 (1979).

\bibitem{Kajantie-78}
  K.~Kajantie, J.~Lindfors and R.~Raitio,
  Phys.\ Lett.\ B {\bf 74}, 384 (1978).

\bibitem{Cleymans-78}
  J.~Cleymans and M.~Kuroda,
  Nucl.\ Phys.\ B {\bf 155}, 480 (1979)
  [Erratum-ibid.\ B {\bf 160}, 510 (1979)].

\bibitem{Lindfors-79}
  J.~Lindfors,
  Phys.\ Scr.\ {\bf 20}, 19 (1979).
  
\bibitem{Lam:1980uc} 
  C.~S.~Lam and W.~K.~Tung,
  Phys.\ Rev.\ D {\bf 21}, 2712 (1980).

\bibitem{Mirkes:1992hu} 
  E.~Mirkes,
  Nucl.\ Phys.\ B {\bf 387}, 3 (1992).
  
\bibitem{Mirkes:1994dp} 
  E.~Mirkes and J.~Ohnemus,
  Phys.\ Rev.\ D {\bf 51}, 4891 (1995)
  [hep-ph/9412289].  

\bibitem{Li:2012wna} 
K.~Melnikov and F.~Petriello,
  Phys.\ Rev.\ D {\bf 74}, 114017 (2006)
  [hep-ph/0609070];
R.~Gavin, Y.~Li, F.~Petriello and S.~Quackenbush,
  Comput.\ Phys.\ Commun.\  {\bf 182}, 2388 (2011)
  [arXiv:1011.3540 [hep-ph]];
  Y.~Li and F.~Petriello,
  Phys.\ Rev.\ D {\bf 86}, 094034 (2012)
  [arXiv:1208.5967 [hep-ph]].
  
\bibitem{Catani:2009sm} 
  S.~Catani, L.~Cieri, G.~Ferrera, D.~de Florian and M.~Grazzini,
  Phys.\ Rev.\ Lett.\  {\bf 103}, 082001 (2009)
  [arXiv:0903.2120 [hep-ph]]; 
S.~Catani and M.~Grazzini,
  Phys.\ Rev.\ Lett.\  {\bf 98}, 222002 (2007)
  [hep-ph/0703012].
  
\bibitem{Peng:2015spa} 
  J.~C.~Peng, W.~C.~Chang, R.~E.~McClellan and O.~Teryaev,
  Phys.\ Lett.\ B {\bf 758}, 384 (2016)
  [arXiv:1511.08932 [hep-ph]].
  
\bibitem{Brandenburg:1993cj} 
  A.~Brandenburg, O.~Nachtmann and E.~Mirkes,
  Z.\ Phys.\ C {\bf 60}, 697 (1993).
  
\bibitem{Brandenburg:1994wf} 
  E.~L.~Berger and S.~J.~Brodsky,
  Phys.\ Rev.\ Lett.\  {\bf 42}, 940 (1979);
  A.~Brandenburg, S.~J.~Brodsky, V.~V.~Khoze and D.~M\"{u}ller,
  Phys.\ Rev.\ Lett.\  {\bf 73}, 939 (1994)
  [hep-ph/9403361]; 
  K.~J.~Eskola, P.~Hoyer, M.~V\"{a}nttinen and R.~Vogt,
  Phys.\ Lett.\ B {\bf 333}, 526 (1994)
  [hep-ph/9404322].
  
\bibitem{Zhou:2009rp} 
  J.~Zhou, F.~Yuan and Z.~T.~Liang,
  Phys.\ Lett.\ B {\bf 678}, 264 (2009)
  [arXiv:0901.3601 [hep-ph]].
 
\bibitem{Konychev:2005iy} 
P.~M.~Nadolsky and C.~P.~Yuan,
  Nucl.\ Phys.\ B {\bf 666}, 3 (2003)
  [hep-ph/0304001];
F.~Landry, R.~Brock, P.~M.~Nadolsky and C.~P.~Yuan,
  Phys.\ Rev.\ D {\bf 67}, 073016 (2003)
  [hep-ph/0212159];
  A.~V.~Konychev and P.~M.~Nadolsky,
  Phys.\ Lett.\ B {\bf 633}, 710 (2006)
  [hep-ph/0506225];
  P.~Sun, J.~Isaacson, C.~P.~Yuan and F.~Yuan,
  arXiv:1406.3073 [hep-ph];
  U.~D'Alesio, M.~G.~Echevarria, S.~Melis and I.~Scimemi,
  J.\ High\ Energy\ Phys.\ {\bf 11}, 098 (2014)
  [arXiv:1407.3311 [hep-ph]];  
  S.~Catani, D.~de Florian, G.~Ferrera and M.~Grazzini,
  J.\ High\ Energy\ Phys.\ {\bf 12}, 047 (2015)
  [arXiv:1507.06937 [hep-ph]].  
  
\bibitem{Bacchetta:2008xw} 
  A.~Bacchetta, D.~Boer, M.~Diehl and P.~J.~Mulders,
  J.\ High\ Energy\ Phys.\  {\bf 08}, 023 (2008)
  [arXiv:0803.0227 [hep-ph]].  

\bibitem{Arnold:2008kf} 
  S.~Arnold, A.~Metz and M.~Schlegel,
  Phys.\ Rev.\ D {\bf 79}, 034005 (2009)
  [arXiv:0809.2262 [hep-ph]];
M.~Boglione and S.~Melis,
  Phys.\ Rev.\ D {\bf 84}, 034038 (2011)
  [arXiv:1103.2084 [hep-ph]]. 

\bibitem{Martin:2009iq} 
  A.~D.~Martin, W.~J.~Stirling, R.~S.~Thorne and G.~Watt,
  Eur.\ Phys.\ J.\ C {\bf 63}, 189 (2009)
  [arXiv:0901.0002 [hep-ph]].

\bibitem{Sutton:1991ay} 
  P.~J.~Sutton, A.~D.~Martin, R.~G.~Roberts and W.~J.~Stirling,
  Phys.\ Rev.\ D {\bf 45}, 2349 (1992).
  
\bibitem{Gluck:1991ey} 
  M.~Gl\"{u}ck, E.~Reya and A.~Vogt,
  Z.\ Phys.\ C {\bf 53}, 651 (1992).

\bibitem{Chiosso:2015naa} 
  M.~Chiosso (COMPASS Collaboration),
  Eur.\ Phys.\ J.\ Web Conf.\  {\bf 85}, 02036 (2015).

\bibitem{Nakano:2016jdm} 
  K.~Nakano (E906/SeaQuest Collaboration),
  Int.\ J.\ Mod.\ Phys.\ Conf.\ Ser.\  {\bf 40}, 1660041 (2016).


\end{thebibliography}
\end{document}